\documentclass[prc,floatfix,nofootinbib,showpacs,showkeys,reprint]{revtex4-1}
\usepackage{graphicx}
\usepackage{amssymb,amsmath,amstext,amsthm,amsfonts}
\usepackage{subfigure}

\newcommand{\dd}{\mathrm{d}}

\begin{document}
\today
\title{Neutrino-Induced Reactions on nuclei}

\author{K. Gallmeister}
\affiliation{Institut f\"ur Theoretische Physik, Johann Wolfgang Goethe-Universit\"at Frankfurt, Frankfurt am Main, Germany}
\author{U. Mosel}
\email[Contact e-mail: ]{mosel@physik.uni-giessen.de}
\affiliation{Institut f\"ur Theoretische Physik, Universit\"at Giessen, Giessen, Germany}
\author{J. Weil}
\affiliation{Frankfurt Institute for Advanced Studies, Frankfurt am Main, Germany}

\begin{abstract}
\begin{description}
\item[Background] Long-baseline experiments such as the planned Deep Underground Neutrino Experiment (DUNE) require theoretical descriptions of the complete event in a neutrino-nucleus reaction. Since nuclear targets are used this requires a good understanding of neutrino-nucleus interactions.
\item[Purpose] Develop a consistent theory and code framework for the description of lepton-nucleus interactions that can be used to describe not only inclusive cross sections, but also the complete final state of the reaction.
\item[Methods] The Giessen-Boltzmann-Uehling-Uhlenbeck (GiBUU) implementation of quantum-kinetic transport theory is used, with improvements in its treatment of the nuclear ground state and of 2p2h interactions. For the latter an empirical structure function from electron scattering data is used as a basis.
\item[Results] Results for electron-induced inclusive cross sections are given as a necessary check for the overall quality of this approach. The calculated neutrino-induced inclusive double-differential cross sections show good agreement data from neutrino- and antineutrino reactions for different neutrino flavors at MiniBooNE and T2K. Inclusive double-differential cross sections for MicroBooNE, NOvA, MINERvA and LBNF/DUNE are given.
\item[Conclusions] Based on the GiBUU model of lepton-nucleus interactions a good theoretical description of inclusive electron-, neutrino- and antineutrino-nucleus data over a wide range of energies, different neutrino flavors and different experiments is now possible. Since no tuning is involved this theory and code should be reliable also for new energy regimes and target masses.
\end{description}
\end{abstract}

\pacs{12.15.-y,25.30.Pt,24.10.Lx}
\keywords{neutrino interactions; long-baseline experiments, neutrino event generators, transport theory}

\maketitle

\section{Introduction}
Unlike in any other nuclear or high-energy physics experiments in neutrino long-baseline experiments the incoming particle's energy is not known; only -- usually quite broad -- energy distributions can be given with some reasonable accuracy. This is a problem because the extraction of the relevant neutrino properties, such as mixing angles, phases and mass-hierarchy, from the measured event rates requires the knowledge of the neutrino energy. The latter must be reconstructed from the observed final state of the reaction.

This task is complicated by the fact that experiments never see the full final state because of experimental limitations (acceptances, thresholds etc). Another complication arises because all presently running and planned experiments at Fermilab and JPARC use nuclear targets, such as C, O or Ar. The energy reconstruction is then possible only with the help of so-called generators \cite{Gallagher:2014pdg}. These generators must be able to describe not only the initial neutrino-nucleon interaction but also the final state interactions of the produced hadrons so that a 'backwards-calculation' from the observed final state to the unknown initial state is possible.

These generators,e.g.\  GENIE \cite{Andreopoulos:2009rq} or NEUT \cite{Hayato:2009zz}, contain, on one hand, a number of purely experimental details, such as target and detector properties and geometrical features, and are thus indispensable. However, they are also being used to distinguish certain elementary interaction modes from each other, for example in background subtractions. In particular at the higher energies, with many open reaction channels, this requires a reliable description of all of them and of their coupling to each other. Thus, generators have to rely on theoretical or phenomenological descriptions of these interactions. Because all running and planned neutrino long-baseline experiments use nuclear targets, such as C, O, Ar or Fe, even in high-energy neutrino experiments, with beam energies in the GeV range, relatively low-energy nuclear physics processes contribute, both to the initial neutrino-nucleus interaction and to the final state interactions (fsi) of outgoing hadrons. It is thus clear that the reliability and predictive power of the generators is the better the more advanced the underlying understanding and implementation of nuclear theory is.

Presently available generators \cite{Andreopoulos:2009rq,Hayato:2009zz,Golan:2012rfa} all rely on free-particle Monte Carlo cascade simulations that are applicable at very high energies but are of only limited applicability in the description of relatively low energy fsi of hadrons inside the target nuclei. A basic feature of nuclei, their binding, is neglected from the outset in these Monte Carlo calculations. Furthermore, the generators often still rely on outdated nuclear and hadron physics and consist of a patchwork of descriptions of different reaction channels without internal consistency \cite{Mosel:2016}.

There is, therefore, now a growing realization in the neutrino long-baseline community that the description of nuclear effects has to be improved. Over the last few years significant theoretical progress has been made in the description of inclusive neutrino-nucleus cross sections. On one hand, state-of-the-art nuclear many-body theories (NMBT) have been used to obtain nuclear ground states and the spectral function of nucleons therein \cite{Benhar:2006wy,Lovato:2015qka}. These can then be used to compute the response of nuclei to electro-weak interactions. While full-scale calculations, the so-called Green's function Monte Carlo calculations \cite{Carlson:2014vla}, are still limited to nuclei lighter than Carbon, more approximate methods have also been used to combine the ground state information with reaction-theoretical models using, e.g., the impulse approximation \cite{Benhar:2005dj}. Nuclear theory has also been used to improve the description of the excited state properties by taking so-called RPA correlations \cite{Kolbe:2003ys,Martini:2016eec} as well as reaction mechanisms that go beyond the impulse approximation into account \cite{Martini:2009uj,Nieves:2011pp}. Other approaches use semi-empirical information obtained from electron scattering \cite{Day:1990mf} to calculate the inclusive electron- and neutrino-induced inclusive cross sections on nuclei \cite{Ivanov:2013saa}. Even though the theoretical progress in our understanding of the inclusive electroweak response of nuclei has been impressive, a drawback is that so far none of these just mentioned theoretical methods can provide the full final state of the $\nu A$ reaction that is needed for the extraction of neutrino properties.

We have, therefore, some years ago started to develop a theory and code framework, called Giessen-Boltzmann-Uehling-Uhlenbeck (GiBUU), that aims to incorporate the 'best' possible information on nuclear interactions into one consistent theory and code that can be used to calculate inclusive cross sections as well as full final-state events; the code can thus be used as an event generator \cite{Buss:2011mx}. 'Best' here does not stand for 'theoretically most advanced', but for 'advanced and practicable to generate both inclusive cross sections and full events'. Different from all other generators GiBUU uses quantum-kinetic transport theory \cite{Kad-Baym:1962}. Transport theory allows one to include important nuclear effects such as binding potentials for hadrons and spectral functions, including their dynamical development. We also require consistency in the sense that the description of all subprocesses, such as, e.g., quasielastic (QE) scattering, pion production, deep inelastic scattering (DIS), and 2p2h interactions, is based on the same ground state.

GiBUU has been used to describe not only $\nu A$ reactions, but also $eA$, $\gamma A$, $\pi A$, $p A$ and even $A A$ reactions and has been widely tested on all of these \cite{Leitner:2009ke,Buss:2011mx}. The description of fsi, which are essentially the same in all of these reactions, has thus extensively been checked against data from all these reactions. We, therefore, restrict ourselves in the present paper to a discussion of results for inclusive cross sections that provide a necessary check for the full model calculations. We do this also because most of the new aspects implemented in a new (February 2016) release of GiBUU have a direct impact on these inclusive reaction rates. In the present paper we give all the details about these new developments. For all other ingredients we refer to previous publications that contain all the details of our description of the various processes in GiBUU \cite{Buss:2011mx}. For neutrino-induced reactions, in particular, details can be found in \cite{Leitner:2006ww,Leitner:2006sp,Leitner:2008ue,Leitner:2009ke,Lalakulich:2012ac,Lalakulich:2012cj} and further references therein.

The main aim of this paper is to demonstrate that now a consistent theoretical description of data from electron-, neutrino- and antineutrino-induced reactions on nuclear targets and in wide energy ranges is possible in a framework that allows at the same time to generate full events. All results shown in this paper can be obtained without any further changes or tunes from the GiBUU 2016 version which can be freely downloaded from \cite{GiBUU}.

\section{New GiBUU Ingredients}
This section briefly discusses the various subprocesses and outlines the aspects that are new in the recent release of GiBUU 2016.

\subsection{Nuclear ground state}
In GiBUU all nucleons are bound in a coordinate- and momentum-dependent potential $U(\mathbf{r},\mathbf{p})$  which is obtained from an analysis of nuclear matter binding properties and $pA$ reactions \cite{Welke:1988zz}; the momentum-dependence is such that a high-momentum nucleon sees a less attractive potential than one with a low momentum. The momentum-distribution is modeled by the local Fermi-gas distribution with $p_F \sim \rho^{1/3}$ . Figure\ 4 in \cite{Alvarez-Ruso:2014bla} shows that the latter reproduces a momentum distribution obtained from NMBT quite well. In particular the significant shift of strength towards lower momentum values, as compared with the distribution of the (global) relativistic Fermi-gas, is reproduced. Missing is the small high-momentum tail that is associated with short range correlations \cite{Benhar:2009hj}.

The preparation of the ground state uses a realistic nuclear density profile, then calculates from an energy-density functional the potential $U$ and, finally, inserts the nucleons into this potential with momenta distributed according to the local Fermi-gas model. The hole spectral function is then given by
\begin{widetext}
\begin{equation}
\mathcal{P}(\mathbf{p},E) = g \int_{nucleus} \dd^3\!r\, f(\mathbf{r},t=0,\mathbf{p}) \Theta(E) \delta\left(E - m^*(\mathbf{r},\mathbf{p}) + \sqrt{\mathbf{p}^2 + {m^*}^2(\mathbf{r},\mathbf{p})}\right)~.
\end{equation}
\end{widetext}
Here $E$ is the removal energy and $f(\mathbf{r},t=0,\mathbf{p})$ is the single-particle Wigner-function in the quasiparticle approximation at time 0; it is the  quantum-kinetic equivalent of the one-particle phase-space distribution. All the potential effects have, for simplicity, been absorbed into a scalar effective mass $m^*(\mathbf{r},\mathbf{p})$ that depends on position $\mathbf{r}$ and momentum $\mathbf{p}$. Due to the integration over the nuclear volume and the $\mathbf{r}$-dependence of the nuclear potential the spectral function of the bound nucleons no longer contains the 'spiky' $\delta$-function in energy, that is typical for the relativistic Fermi gas; instead, it is smeared out. The spectral function is thus, also in its energy-dependence, similar to the realistic one obtained from NMBT \cite{Benhar:2006wy}.

In this method to prepare the ground state the energy of the Fermi surface
\begin{equation}
E_F = \frac{p_F^2(r)}{2M} + U[\rho(r),p_F(r)] \qquad {\rm with} \quad p_F(r) \sim \rho^{1/3}(r)
\end{equation}
is not constant throughout the nuclear volume. Towards the nuclear surface this energy usually rises, thus distorting the energy distribution and making the target nucleus unstable.

In GiBUU 2016 we have now cured this problem by fixing the value of $E_F$ from the outset\footnote{The default value is $E_F = - 8$ MeV, but this value could be changed from nucleus to nucleus.}. This is achieved by calculating the potential for a conventional, realistic Woods-Saxon density distribution. Then, by keeping the functional form of the potential and the value of the Fermi-energy fixed, a nonlinear equation for the density is solved by iteration. The method is similar to the one used in \cite{Alberico:1997jg}.

\subsection{QE scattering, pion production, and DIS}
\begin{itemize}

\item \emph{True QE scattering}, i.e.\ QE scattering on one nucleon, is described as outlined in Ref.\ \cite{Leitner:2006ww}. The axial form factor is assumed to be of dipole form with an axial mass $M_A = 1.03$ GeV. Both the initial and the final state of the nucleon experience the same nuclear potential, but at different momenta. The final state potential of the outgoing nucleon is less attractive than that of the bound ground state nucleon (see Figure\ 7.1 in \cite{Leitner:2009zz}).

\item \emph{Pion production} has been described in detail in Refs.\ \cite{Lalakulich:2010ss,Lalakulich:2012cj}; for the $\Delta$ resonance energy regime the very same theory has been used by the authors of Ref.\ \cite{Hernandez:2013jka}. It proceeds through nucleon resonances with invariant masses less than 2 GeV. The $\Delta$ resonance dominates the resonance pion production; its width is either taken to be that of a free $\Delta$, or that of a collision-broadened one as parameterized by Oset and collaborators \cite{Oset:1987re}. The background terms for electron-induced reactions are obtained by subtracting the calculated resonant contributions from results of the MAID analysis \cite{Tiator:2009mt}. For neutrinos this background contribution is either taken from an effective field theory model, as described in \cite{Lalakulich:2010ss}, or as a parameterized multiple of the vector coupling background, as described in \cite{Leitner:2006ww}. The coupling to higher lying resonances is determined by PCAC with a dipole form factor with $M_A = 1$ GeV. While for electrons also a non-resonant $2\pi$ background amplitude has been implemented, there is no such contribution for the neutrino-induced reactions because of the absence of any experimental information on that background.

\item \emph{DIS} is handled by the \textsc{pythia}, v. 6.4, code \cite{Sjostrand:2006za}; it sets in at invariant masses of about 2 GeV.  The binding of nucleons in a potential poses a problem that becomes essential mainly close to particle production thresholds, as explained in more detail in \cite{Buss:2011mx,Lalakulich:2012gm}.  For electrons \textsc{pythia}, v.6.4,  contains an elaborate model implementation that also includes, e.g., VMD contributions to particle production at the higher energies. No such processes are taken into account in \textsc{pythia} for neutrino-induced reactions.

For reasons of consistency we want to describe both processes, with electrons and those with neutrinos, within the same theory and code and thus use the perturbative 'neutrino-machinery' also for electrons with the properly modified coupling constants. DIS processes become active for larger momentum transfers $Q^2 \gtrsim 1$ GeV$^2$. It, therefore,  has to be made sure that the DIS cross section on a nucleon becomes 0 for $Q^2 \to 0$.  Furthermore, at low $Q^2$ partons have to be screened. For electron-induced reactions both of these effects can be achieved with an ansatz for the $\gamma^*$-nucleon interaction of the form \cite{Friberg:2000nx}
\begin{equation}    \label{DIS}
\sigma_{\rm DIS}^{\gamma\!*p} = \left(\frac{Q^2}{Q^2 + m_v^2}\right)^n \frac{4 \pi^2 \alpha }{Q^2} F_2(x,Q^2)  ~,
\end{equation}
with the exponent $n=2$ and $m_v$ being a mass of the order of hadronic mass scale such as the $\rho$ meson mass; $\alpha$ is the electromagnetic coupling constant and $F_2$ the structure function. Due to the absence of the photon propagator, for neutrinos the requirement of a finite cross section is not necessary and only one such cut-off factor is needed. In GiBUU we, therefore, use for neutrinos the exponent $n=1$. The presence of the attenuation factor in Eq.\ (\ref{DIS}) is a new feature in GiBUU 2016. It lowers the DIS cross section at low $Q^2$, and consequently also the total DIS contribution, by a noticeable amount. An observable consequence is a lowering of the pion production cross section.

\end{itemize}

\subsection{2p2h Interactions}
It was realized early on that so-called 2p2h excitations, in which the incoming neutrino interacts with a correlated pair of nucleons, can contribute to QE-like events in Cerenkov detectors \cite{Delorme:1985ps}. This is so because their experimental signature (1 $\mu^-, 0 \pi$) is indistinguishable from that of true QE scattering\footnote{Also events in which pions were initially produced, but later reabsorbed, contribute to QE-like events.}. Martini \textit{\textit{et al.}}\ \cite{Martini:2009uj} rediscovered this as an explanation for high QE-like cross section observed in the MiniBooNE. These authors could indeed show that a calculation based on a free local Fermi gas with RPA excitations and 2p2h interactions explains the MiniBooNE data, both for neutrinos and antineutrinos\cite{Martini:2010ex,Martini:2011wp}. Subsequently, this same result was also obtained by Nieves and collaborators \cite{Nieves:2011pp,Nieves:2011yp,Nieves:2013fr}

We had shown in Ref.\ \cite{Lalakulich:2012ac} that the observed MiniBooNE double-differential cross sections could be explained quite well also in a model in which the hadronic 2p2h tensor was parameterized in a very simplistic way. Sensitivities to details of the hadronic tensor were obviously wiped out by the flux average. However, this procedure was unsatisfactory because it required arbitrary, new strength parameters for neutrinos and antineutrinos and for electrons.

These shortcomings, that limit the predictive power significantly, have now been overcome in GiBUU 2016 \cite{GiBUU}. We start from the assumption that the dominant 2p2h contributions are transverse. This assumption finds its justification in microscopic studies of 2p2h processes \cite{Martini:2009uj,Megias:2016lke,Simo:2016ikv}. For all calculations, both for electrons and for neutrinos, we neglect the lepton masses in the 2p2h component, but do take them into account in all the other processes.

\subsubsection{Electrons}
The 2p2h contribution to the cross section for scattering of electrons is in general given by
\begin{equation} \label{s-em}
\frac{\dd^2\sigma^{2p2h}}{d\Omega dE'} = \frac{4 \alpha^2 }{Q^4} E'^2 \left(2W_1^e \sin^2\frac{\theta}{2} + W_2^e \cos^2 \frac{\theta}{2} \right) ~,
\end{equation}
where $W_1$ and $W_2$ are structure functions for the 2p2h process and $E'$ is the outgoing lepton's energy.  For a purely transverse interaction we have
$
W_2^e = \frac{Q^2}{\mathbf{q}^2} W_1^e   ~,
$
so that the cross section becomes
\begin{equation} \label{s-em1}
\frac{\dd^2\sigma^{2p2h}}{d\Omega dE'} = \frac{ 8 \alpha^2 }{Q^4} E'^2 \cos^2 \frac{\theta}{2} \left(\frac{Q^2}{2 \mathbf{q}^2} + \tan^2\frac{\theta}{2}\right) W_1^e(Q^2,\omega) ~.
\end{equation}
Thus in this case only one structure function, $W_1^e(Q^2,\omega)$, that depends on the squared four-momentum transfer $Q^2 = \mathbf{q}^2 - \omega^2$ and energy transfer $\omega$, determines the cross section.

\begin{figure}[h]
\centering
\includegraphics[width=\linewidth]{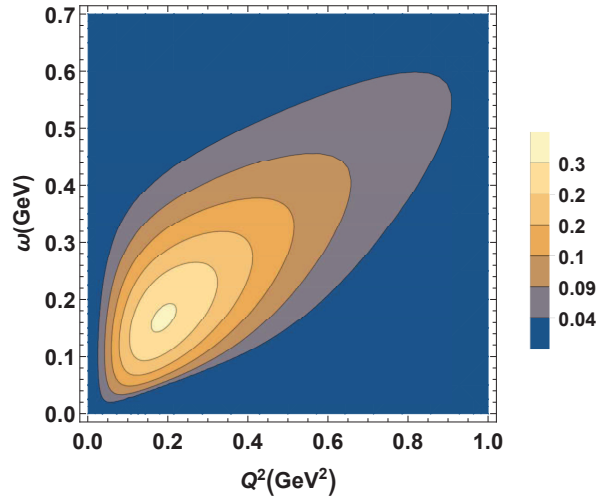}
\caption{Structure function $W_1^{\rm MEC}(Q^2,\omega)$ for $^{12}$C in the parametrization of Bosted \cite{Bosted:2012qc} and Christy \cite{Christy:2015}, in units of 1/GeV.}
\label{fig:W1}
\end{figure}

This part of the total inclusive cross section, encoded in the function $W_1^e(Q^2,\omega)$, has been determined by Bosted and Mamyan \cite{Bosted:2012qc}\footnote{Ref.\ \cite{Bosted:2012qc} contains a number of essential misprints. Based on the code that was used to obtain the structure functions these have been corrected for the present study}. These authors analyzed electron inclusive data on different target nuclei over a wide range of $Q^2$ from 0 to 10 GeV$^2$ and invariant masses between 0.9 and 3.0 GeV.  The QE component in this analysis was obtained from the scaling model and an inelastic contribution was modeled by suitable parametrizations of the structure functions. A so-called meson exchange current (MEC) term was determined to describe all remaining effects. We use this $W_1^{\rm MEC}$ from \cite{Bosted:2012qc}, with a recent improvement at small $Q^2$ by Christy \cite{Christy:2015}, in GiBUU 2016. Figure\ \ref{fig:W1} shows the ($Q^2,\omega$) dependence of this structure function.
The structure function $W_1^e = W_1^{\rm MEC}$ contains all effects beyond true QE and inelastic scattering. In particular, it includes effects of meson exchange currents (genuine 2p2h), short-range correlations and RPA correlations. A feature of the particular parametrization determined by Christy \cite{Christy:2015} is that the structure function vanishes for $Q^2=0$. This is in contrast to earlier theoretical descriptions of photonuclear reactions \cite{Carrasco:1989vq,Effenberger:1996im,Gil:1997bm} that give some strength also at $Q^2=0$ for higher energy transfers starting roughly at the peak of the $\Delta$ resonance. We will come back to this point later in the discussion in Sect.\ \ref{s:2p2hsize}.

\subsubsection{Neutrinos}
While for electrons the 2p2h contribution, under the assumption of being purely transverse, could be described by only one structure function, for neutrinos one more structure function is needed, $W_3$. The cross section is then given by
\begin{eqnarray}  \label{LPnu}
\frac{\dd^2\sigma^{2p2h}}{d\Omega dE'}
&=& \frac{G^2}{2 \pi^2} E'^2 \cos^2 \frac{\theta}{2} \,\left[2W_1^\nu \left(\frac{Q^2}{2\mathbf{q}^2} + \tan^2\frac{\theta}{2}  \right) \right.  \nonumber \\
& & \mbox{}\left. \mp W_3^\nu \frac{E + E'}{M} \tan^2\frac{\theta}{2}\right] ~.
\end{eqnarray}
In Ref.\ \cite{O'Connell:1972zz} the authors have shown that for non-relativistic systems $W_1^\nu$ can be directly related to $W_1^e$
\begin{equation}    \label{Wone}
W_1^\nu = \left[1 + \left(\frac{\mathbf{q}}{\omega}\frac{G_A}{G_M}\right)^2 \right] 2  \,(\mathcal{T} + 1)\, W_1^e ~,
\end{equation}
and $G_M$ and $G_A$ are the magnetic and axial form factors, respectively, and $\mathcal{T}$ is the isospin of the target. For both form factors we use dipole forms with the vector and the axial mass cut-off parameters being 0.84 and 1.03 GeV, respectively.

Furthermore, the longitudinal response can be neglected in the V-A interference terms \cite{O'Connell:1972zz}; this directly relates the structure function $W_3$ to $W_1$. This property is also exploited by Martini \textit{et al.}\ \cite{Martini:2009uj}. The transverse part of their cross section is given by
\begin{eqnarray}
\frac{\dd^2\sigma^{2p2h}}{d\Omega dE'}
&=& \frac{G^2}{2 \pi^2} E'^2 \left[2\left(G_M^2\frac{\omega^2}{\mathbf{q}^2} + G_A^2\right) \left(\frac{Q^2}{2\mathbf{q}^2} + \tan^2\frac{\theta}{2} \right)\right. \nonumber \\
& & \mbox{} \left. \mp 2 G_A G_M \frac{E + E'}{M_N} \tan^2\frac{\theta}{2} \right] R_{\sigma \tau}^\nu(T) ~,
\end{eqnarray}
where $R_{\sigma \tau}^\nu(T)$ is the transverse spin-isospin response. Note that the same response $R_{\sigma \tau}(T)$ appears both in the direct and the interference term, so that $W_1 \sim W_3$. By comparison with \cite{O'Connell:1972zz} we obtain
\begin{eqnarray}    \label{sigma-nu}
W_1^\nu
&=& \left(G_M^2\frac{\omega^2}{\mathbf{q}^2} + G_A^2\right)  R^\nu_{\sigma \tau}(T)    \nonumber \\
&=& \left(G_M^2\frac{\omega^2}{\mathbf{q}^2} + G_A^2\right) \frac{1}{2 G_A G_M} W_3^\nu ~.
\end{eqnarray}

Even though we have denoted the cross section by the superscript 2p2h, we stress again that this contains a mixture of effects from meson-exchange interactions and short- and long-range 2p2h interactions, as well as possible RPA effects.

\subsubsection{A-dependence}     \label{A-dep}
The dependence of the 2p2h cross section on mass number $A$ is obtained by assuming a short-range interaction so that the two nucleons are localized at the same location with momenta taken randomly from the Fermi-sea. The interaction of such pairs of nucleons at the same location is proportional to the average nuclear density $\int \dd^3\!r \,\rho^2(r) = A \langle \rho(r)\rangle$. For large $A$ it increases linearly with $A$. For smaller $A$ the importance of the nuclear surface increases relative to the volume \cite{Mosel:2016uge}. The number then drops below a linear dependence, in agreement with results found in a more sophisticated microscopical calculation \cite{Vanhalst:2014cqa,Colle:2015ena}. In GiBUU the isospin composition of pairs is chosen randomly, such that, e.g., for $^{12}$C the probability for pp pairs vs.\ pn pairs is $Z(Z-1)/(2ZN) = 5/12 \approx 0.42$ and for $^{40}$Ar it amounts to $17/44 \approx 0.39$.

\section{Inclusive Electron Cross sections}
In this section we discuss the inclusive electron cross sections obtained with the model described in the previous section, with an emphasis on 2p2h excitations. Since the structure function $W_1^e$ has been fitted to data the comparison with data here serves as a consistency check. It is checked that even though Bosted \textit{et al.}\ \cite{Bosted:2012qc} used different QE- and DIS-components than in the present model, the GiBUU calculations still give cross sections that agree with the data. Furthermore, the numerical implementation is tested for correctness, since the electron- and neutrino cross sections are calculated from one and the same part of the code and not some separate module.

\begin{figure}[ht]
\centering
\includegraphics[width=0.65\linewidth,angle=-90]{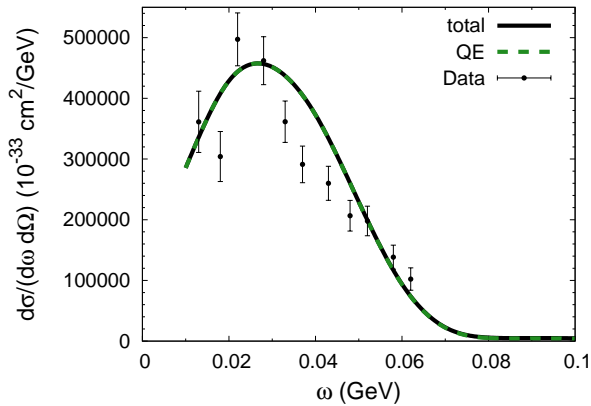}
\caption{Inclusive cross section for scattering of electrons on carbon at 240 MeV and 36 Deg ($Q^2 = 0.02$ GeV$^2$ at the QE peak). In this case the total cross section equals that of the QE-scattering process alone; other contributions are negligible. The data were obtained from the Quasielastic Electron Nucleus Scattering Archive \cite{Benhar:2006er}. }
\label{fig:eC240-36}
\end{figure}

Figure \ref{fig:eC240-36} shows results for a very small $Q^2 = 0.02$ GeV$^2$ where the applicability of the impulse approximation becomes doubtful. Here the calculated cross section is dominated by true QE scattering; the 2p2h component, as well as other reaction processes, do not show up on the scale of this figure. The calculated response is slightly (by about 10 MeV) shifted towards higher energy transfers. This small shift is well within the accuracies of the method. The peak position is sensitive to the momentum dependence of the mean field potential \cite{Brieva:1977dv,Rosenfelder:1978qt,Ankowski:2014yfa} and the momentum-dependence encoded in GiBUU is not very detailed at such small momenta.

\begin{figure}[h]
\centering
\includegraphics[width=0.65\linewidth,angle=-90]{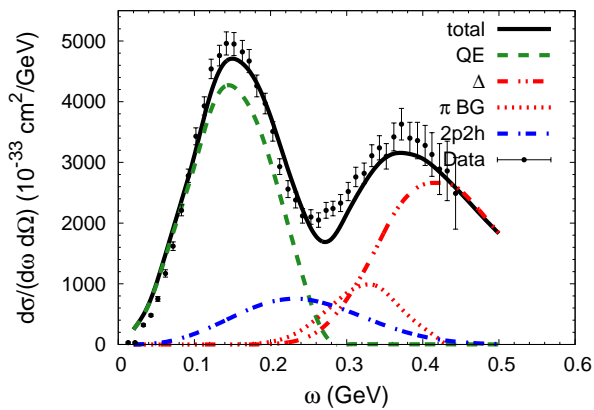}
\caption{Inclusive cross section for scattering of electrons on carbon at 560 MeV and 60 Deg ($Q^2 = 0.24$ GeV$^2$ at the QE peak), obtained with a free $\Delta$ spectral function. The leftmost dashed curve gives the contribution from true QE scattering, the dash-dotted curve that from 2p-2h processes, the dashed-dotted-dotted curve that from $\Delta$ excitation and the dotted curve that from pion background terms. Data from \cite{Benhar:2006er}. }
\label{fig:eC560-60}
\end{figure}

Results for a larger $Q^2 = 0.24$ GeV$^2$ are shown in Figure\ \ref{fig:eC560-60}. Now other reaction channels besides just true QE contribute. Under the QE peak the 2p2h contribution amounts to about 10\%; it peaks in the dip region between the QE-peak and the $\Delta$-peak. In the dip region there is also already a significant contribution not only from the $\Delta$ resonance, but also from non-resonant background in pion production.

\begin{figure}[h]
\centering
\includegraphics[width=0.65\linewidth,angle=-90]{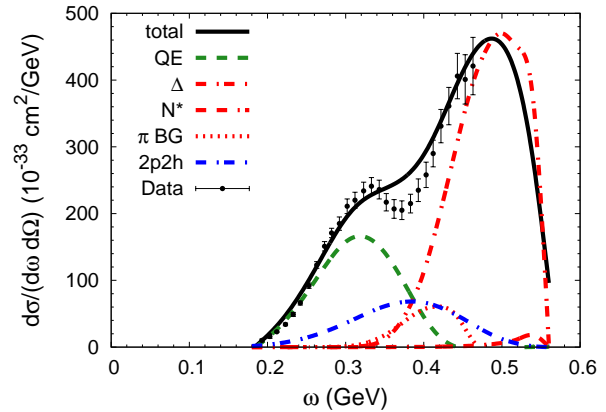}
\caption{Inclusive cross section for scattering of electrons on carbon at 560 MeV and 145 Deg ($Q^2 = 0.55$ GeV$^2$ at the QE peak), obtained with a free $\Delta$ spectral function. The various contributions are indicated in the figure; they are the same as in Figure\ \ref{fig:eC560-60}. Data from \cite{Benhar:2006er}. }
\label{fig:eC560-145}
\end{figure}

Finally, in Figure\ \ref{fig:eC560-145} we show results for an even larger $Q^2 = 0.55$ GeV$^2$. 2p2h processes again contribute about 10\% under the QE-peak. The dip-region is now somewhat overestimated, this could be due to an overestimation of the background terms for pion production at this larger $Q^2$. At very large $\omega$ now also a small contribution from higher nucleon resonances shows up, but the region beyond the QE peak is still dominated by the $\Delta$ and the background terms. Interesting here is the behavior of the total cross section above $\omega \approx 0.5$ GeV. Here the total cross section (black solid line) is smaller than the $\Delta$ contribution (red dash-dot-dot line) reflecting a negative interference between resonance and background amplitudes.

\section{Inclusive Neutrino Cross Sections}
The results for electrons discussed in the last section constitute a consistency check for the numerical implementation in GiBUU.  We now discuss neutrino inclusive cross sections calculated using the nuclear structure functions $W_1^\nu$ of Eqs.\ (\ref{Wone}) and $W_3^\nu$ of Eq.\ (\ref{sigma-nu}). All of these results were obtained with the isospin factor $\mathcal{T} = 1$ in Eq.\ (\ref{Wone}).

\subsection{MiniBooNE results}
We start with a discussion of the MiniBooNE results \cite{AguilarArevalo:2010zc,Aguilar-Arevalo:2013dva} for neutrinos and antineutrinos because these are still the only double-differential data available for a wide range of muon angles and energies.

\subsubsection{Neutrinos}
\begin{figure*}[ht]
\centering
\includegraphics[width=\linewidth]{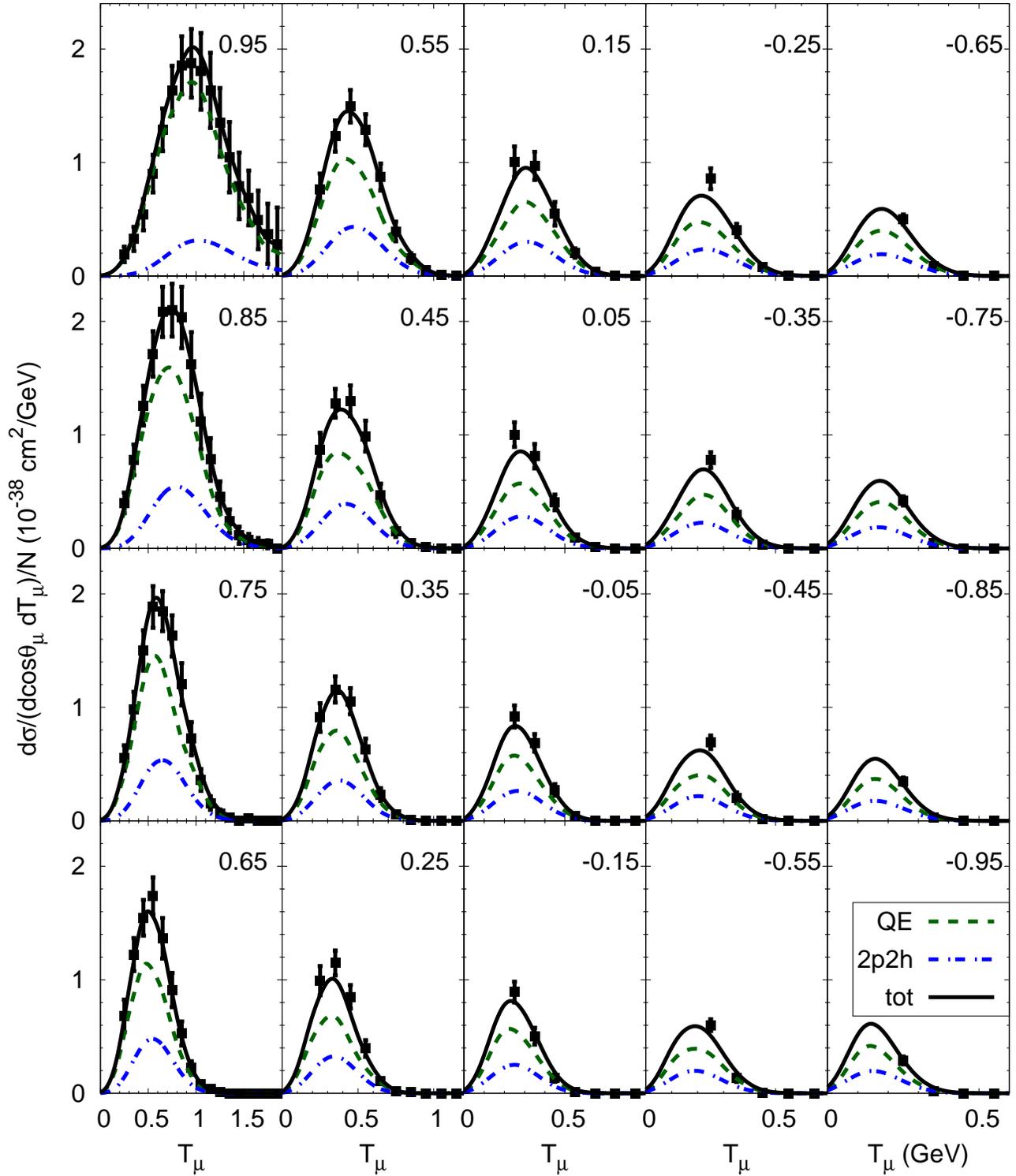}
\caption{Double differential cross section per neutron for QE + 2p2h neutrino events with a $^{12}$C target at MiniBooNE. The numbers in the upper right corner of the individual frames give the cosine of the scattering angle, $T_\mu$ is the outgoing muon's kinetic energy in GeV. Note that the energy scale changes between the first and the following columns. The lowest, dashed-dotted blue curve gives the 2p2h contribution, the middle, green dashed curve that of QE scattering and the topmost solid black curve gives the total. Data are taken from \cite{AguilarArevalo:2010zc}.}
\label{fig:MB-dd+QE-nu}
\end{figure*}

Figure\ \ref{fig:MB-dd+QE-nu} shows the neutrino results for the QE cross section as obtained by MiniBooNE by subtracting the so-called stuck-pion events\footnote{Stuck pion events are those in which initially a pion or $\Delta$ resonance were produced and then, later on, reabsorbed.} from their inclusive '0 pion' data.
\begin{figure}
\includegraphics[width=0.7\linewidth,angle=-90]{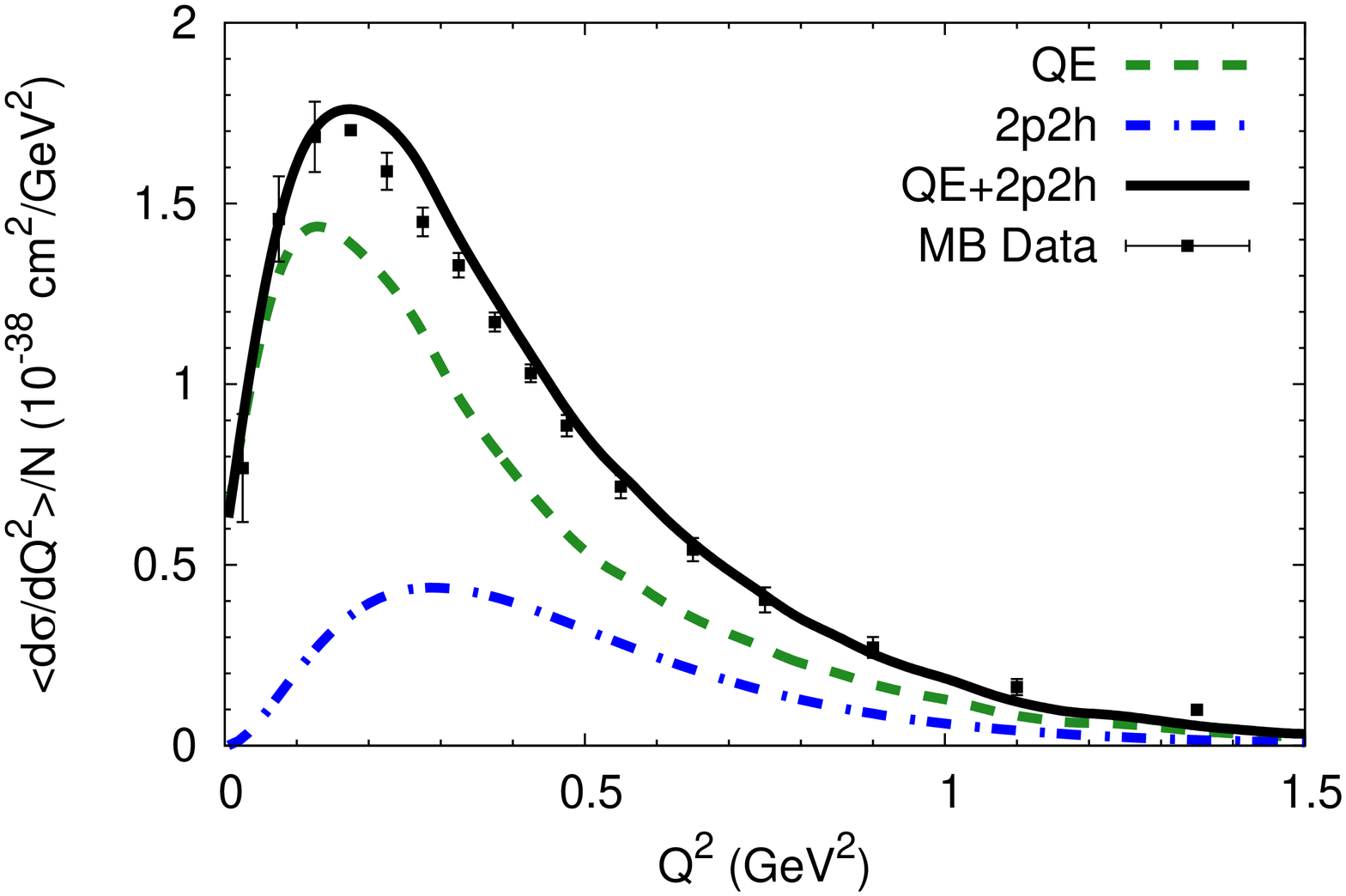}
\caption{$Q^2$ distribution per neutron for the MiniBooNE QE + 2p2h neutrino events on a C target. Data are taken from \cite{AguilarArevalo:2010zc}.}
\label{fig:MB-dd+Q2-nu}
\end{figure}
The agreement is excellent over the full range of energies and angles and of the same quality as the results obtained by Martini \textit{et al.}\  \cite{Martini:2011wp}. There has been no readjustment of the published flux whereas in the work of Nieves \textit{et al.}, who obtained a similar agreement, the experimental cross sections were scaled down by about 10\% \cite{Nieves:2011yp}.

The relative weight of 2p2h processes increases with increasing angle, as a consequence of the transverse nature of this process. The comparison of the $Q^2$ distribution is shown in Figure\ \ref{fig:MB-dd+Q2-nu}. Again, the agreement is quite good, with the calculated values being slightly higher than the experimental values around 0.2 GeV$^2$. However, contrary to the double-differential cross sections shown in Figure \ref{fig:MB-dd+QE-nu}, the experimental cross section shown here is plotted versus a reconstructed quantity which brings some uncertainties with it \cite{Lalakulich:2012hs}.

\subsubsection{Antineutrinos}
\begin{figure}[!ht]
\centering
\includegraphics[width=\linewidth]{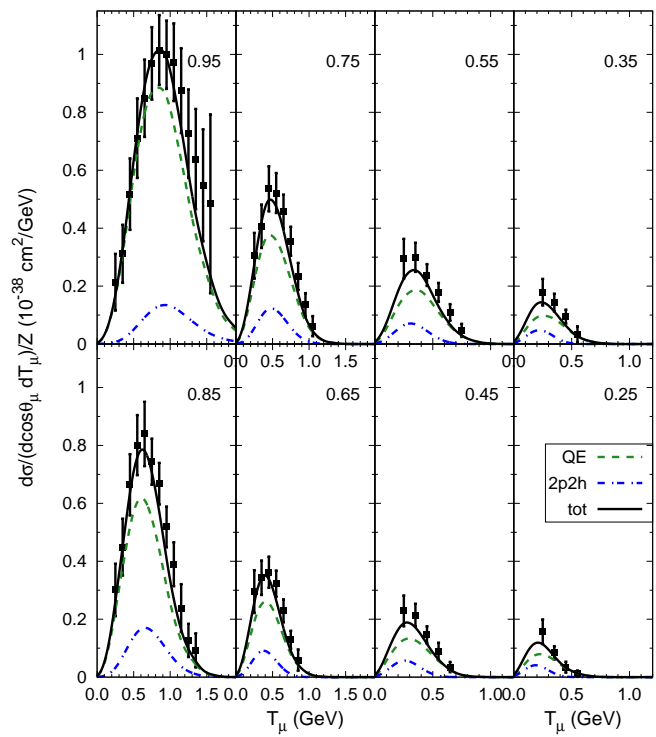}
\caption{Forward angle double differential cross section per proton for QE + 2p2h events in the antineutrino beam in the MiniBooNE with a $^{12}$C target. The numbers in the upper right corner give the cosine of the muon scattering angle while $T_\mu$ is the muon's kinetic energy in GeV. The lowest, dashed-dotted blue curve gives the 2p2h contribution, the middle, dashed green curve the true QE contribution and the uppermost solid black curve the sum. Data are taken from \cite{Aguilar-Arevalo:2013dva}.}
\label{fig:MB-barnu}
\end{figure}

\begin{figure}[h]
\centering
\includegraphics[width=\linewidth]{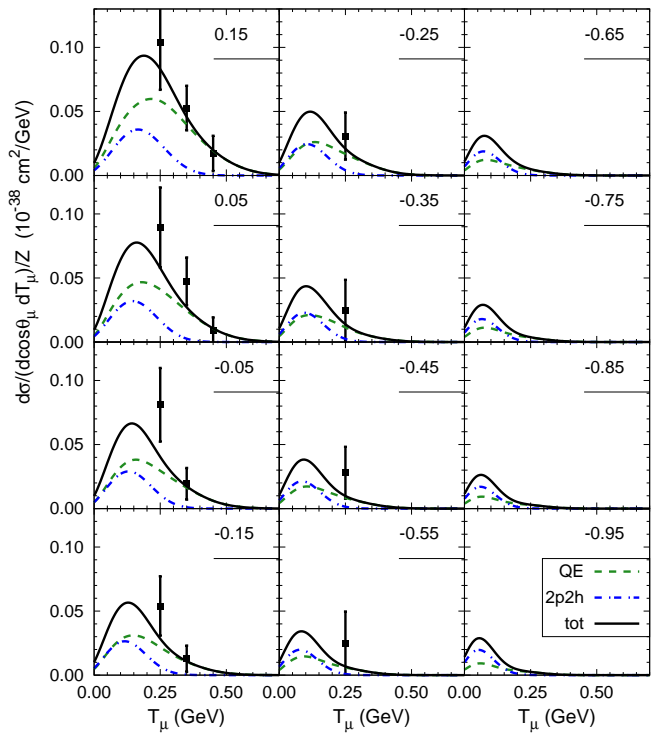}
\caption{Same as Figure\ \ref{fig:MB-barnu} in the backward region with a stretched cross sections scale. Data from \cite{Aguilar-Arevalo:2013dva}.}
\label{fig:MB-barnu-scaled}
\end{figure}

Using the same structure functions also for antineutrinos yields the results shown in Figure\ \ref{fig:MB-barnu}.
The cross section is now more strongly forward peaked than for neutrinos due to the weakening of the transverse 2p2h component because of the opposite sign in front of the $W_3$ structure function. A closer inspection of these backwards angles in Figure\ \ref{fig:MB-barnu-scaled} shows that there the 2p2h cross section becomes comparable to that for the true QE process. At the most backward angle of $\cos \theta = -0.95$ the 2p2h cross section is even dominant.
\begin{figure}[h!]
\centering
\includegraphics[width=0.7\linewidth,angle=-90]{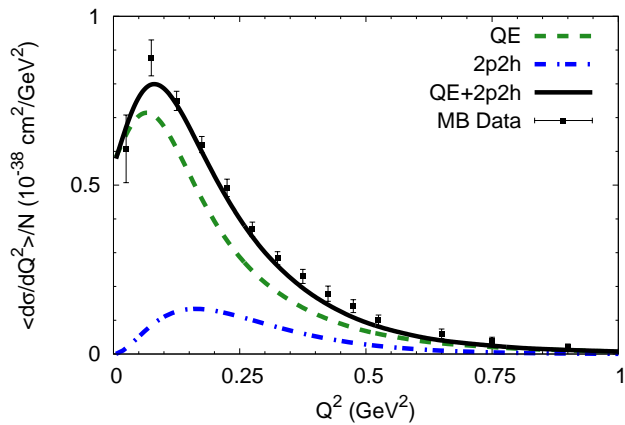}
\caption{$Q^2$ distribution of QE + 2p2h events for the MiniBooNE antineutrino flux for a C target. Curves as in Figure\ \ref{fig:MB-dd+Q2-nu}.  Data are taken from \cite{Aguilar-Arevalo:2013dva}. }
\label{fig:MB-barnu-dsigmadQ2}
\end{figure}

It is also noticeable that for the most forward angular bin there is a discrepancy at the higher muon energies above about 1 GeV where the calculated cross section falls short of the experimental data. Exactly the same discrepancy also shows up in the results of Nieves \textit{et al.}\ \cite{Nieves:2013fr}. A related discrepancy is observed in the calculation of the $Q^2$ distribution shown in Figure\ \ref{fig:MB-barnu-dsigmadQ2}. Here the peak is somewhat underestimated; the same behavior also shows up in the results of Martini \textit{et al.}\ \cite{Martini:2013sha}. Considering that the published absolute flux uncertainty \cite{Aguilar-Arevalo:2013dva} is about 13\% we consider this again quite good agreement.
Thus two discrepancies show up in independent calculations, the underestimate of the calculated cross section at high $T_\mu > 1$ GeV for the very forward direction (small $Q^2$) and the underestimate at $Q^2 \approx 0.1$ GeV$^2$. This leads us to conjecture that both of these are related and are due to an underestimate of the flux in its high-energy tail at energies above about 1 GeV.

\subsection{T2K near detector results}
\begin{figure}[ht]
\begin{center}
\subfigure{
\includegraphics[width=0.8\linewidth,angle=0]{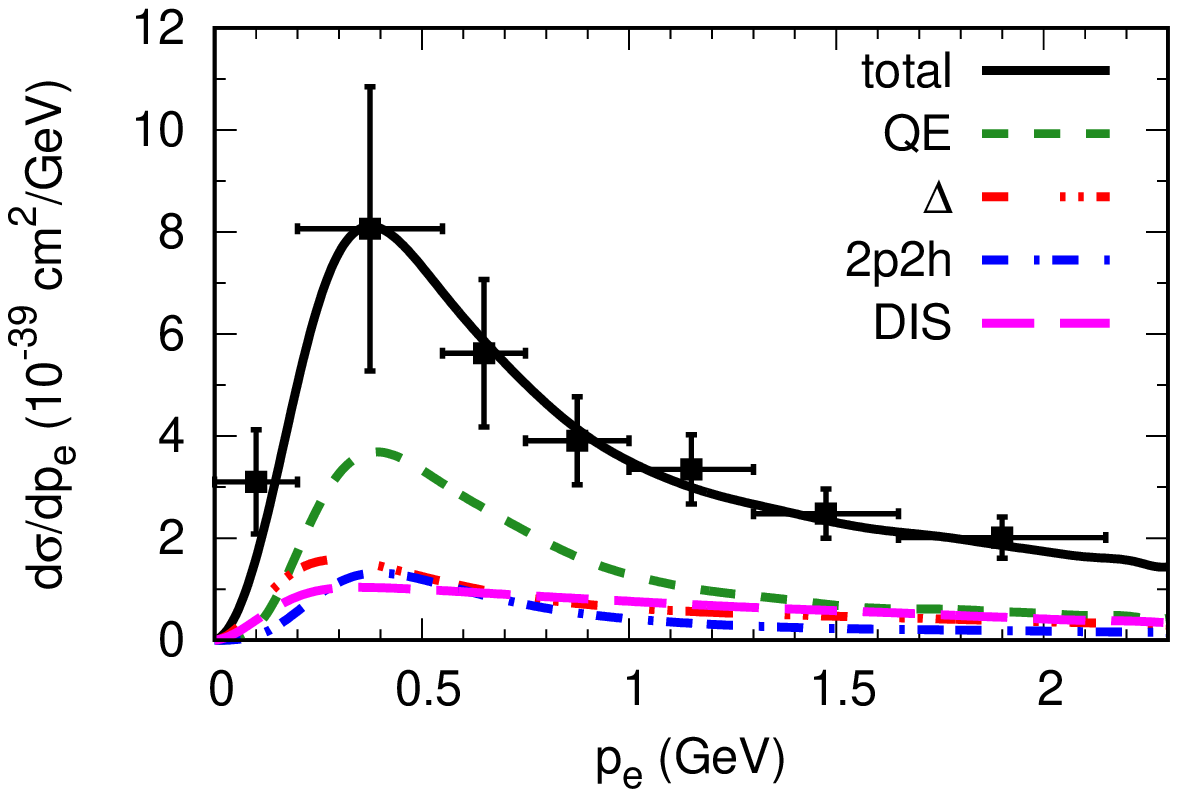}
}
\subfigure{
\includegraphics[width=0.8\linewidth,angle=0]{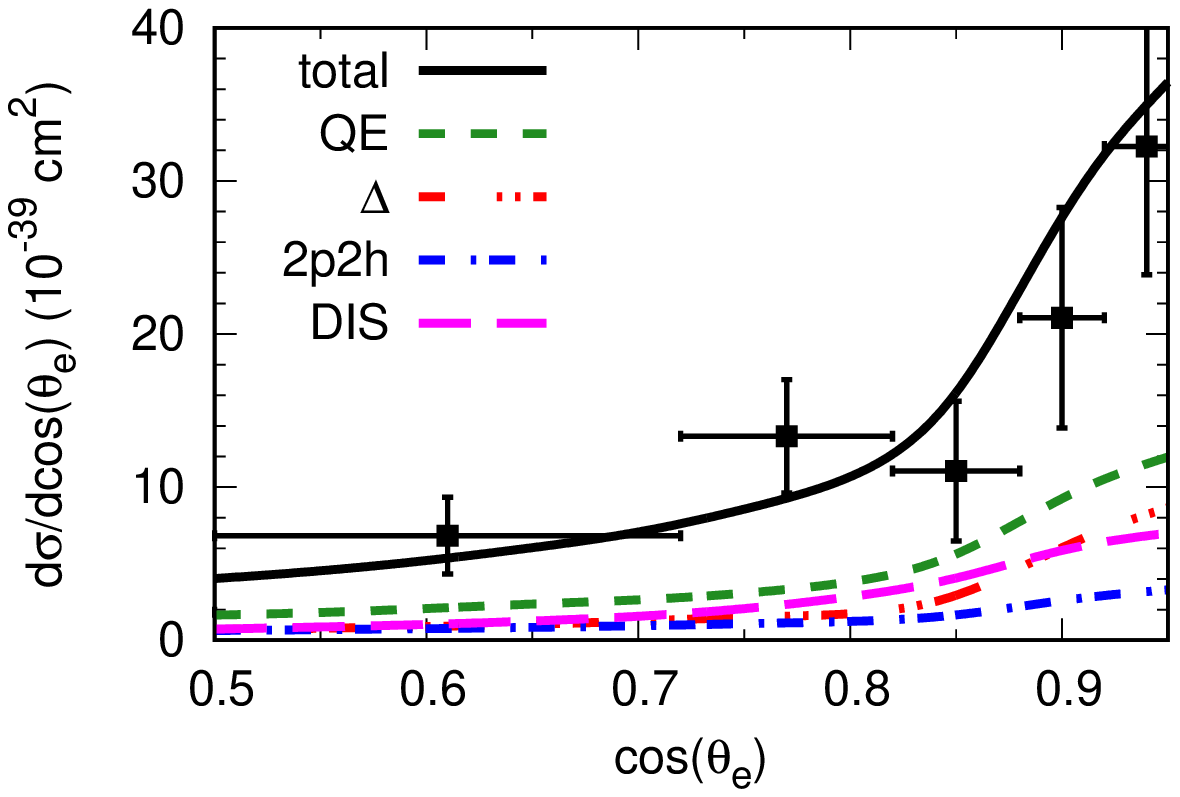}
}
\end{center}
\caption{Momentum (top panel) and angular distribution (bottom panel) of fully inclusive events for an electron-neutrino beam on a C target in the T2K near detector. The solid curve gives the sum of all contributions; the contributions of some dominant reaction channels are explicitly indicated in the figure: QE (green, dashed), $\Delta$ excitation (red, dash-dot-dotted), 2p2h (blue, dash-dotted) and DIS (magenta, long-dashed). The data are taken from \cite{Abe:2014agb}.}
\label{fig:T2K_ND_p-ecost}
\end{figure}

In order to subject these results and the underlying model to an independent test we now show results obtained at the near detector of the experiment T2K with an electron-neutrino beam. The flux distribution of this experiment is similar to that of MiniBooNE, but slightly narrower in energy. In Figure\ \ref{fig:T2K_ND_p-ecost} we show the momentum- and angular-distributions for the fully inclusive cross section for the interaction of electron neutrinos with $^{12}$C in comparison with the experimental values \cite{Abe:2014agb}. The agreement is very good and of the same quality as in recent RPA calculations \cite{Martini:2016eec}. Integrated over all angles the $\Delta$ contribution is comparable to the 2p2h component even at this relatively low energy.

Since the electron-neutrino flux in the T2K experiment contains a high-energy tail it is interesting to look explicitly also at the DIS contribution. This is given by the long-dashed (magenta) curve in Figure\ \ref{fig:T2K_ND_p-ecost}. At forward angles and small outgoing electron momenta DIS is seen to contribute as much to the total inclusive cross section as the $\Delta$ resonance with a similar dependence on $p_e$ and $\cos \theta_e$, respectively. The agreement is as good as that obtained in \cite{Martini:2016eec} where, however, the DIS component is absent. A closer inspection of that result shows that while the QE and 2p2h components are similar in magnitude the 1$\pi$ incoherent component obtained there is significantly larger than that obtained here. This is probably a consequence of the missing pion absorption in the calculations of Ref.\ \cite{Martini:2016eec}.
\begin{figure}[ht!]
\centering
\includegraphics[width=1.\linewidth]{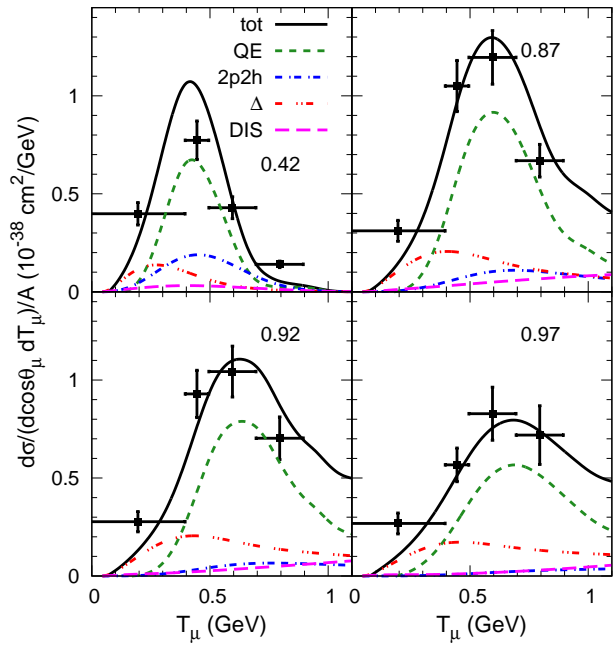}
\caption{Inclusive double-differential cross section per nucleon for a C target with the muon-neutrino beam in the T2K near detector. The numbers in the individual plots give the $\cos \theta$ of the outgoing muon. The solid curve gives the sum of all contributions; the contributions of some dominant reaction channels are explicitly indicated in the figure. Data are taken from \cite{Abe:2013jth}}
\label{fig:T2K_ND}
\end{figure}

A similarly good agreement is reached for the inclusive double-differential cross section, shown in Figure\ \ref{fig:T2K_ND} for a muon-neutrino beam at the near detector of T2K \cite{Abe:2013jth}. The 2p2h contribution is of only minor importance for most of the angles, except for the largest one ($\cos \theta = 0.42$). There it amounts to about 20\% of the total; at the other angles it is significantly smaller. DIS gives a small contribution at all angles; except for the largest one it is very close to that of 2p2h processes.

\subsection{MicroBooNE results}
\begin{figure*}[ht]
	\centering
	\includegraphics[width=\linewidth]{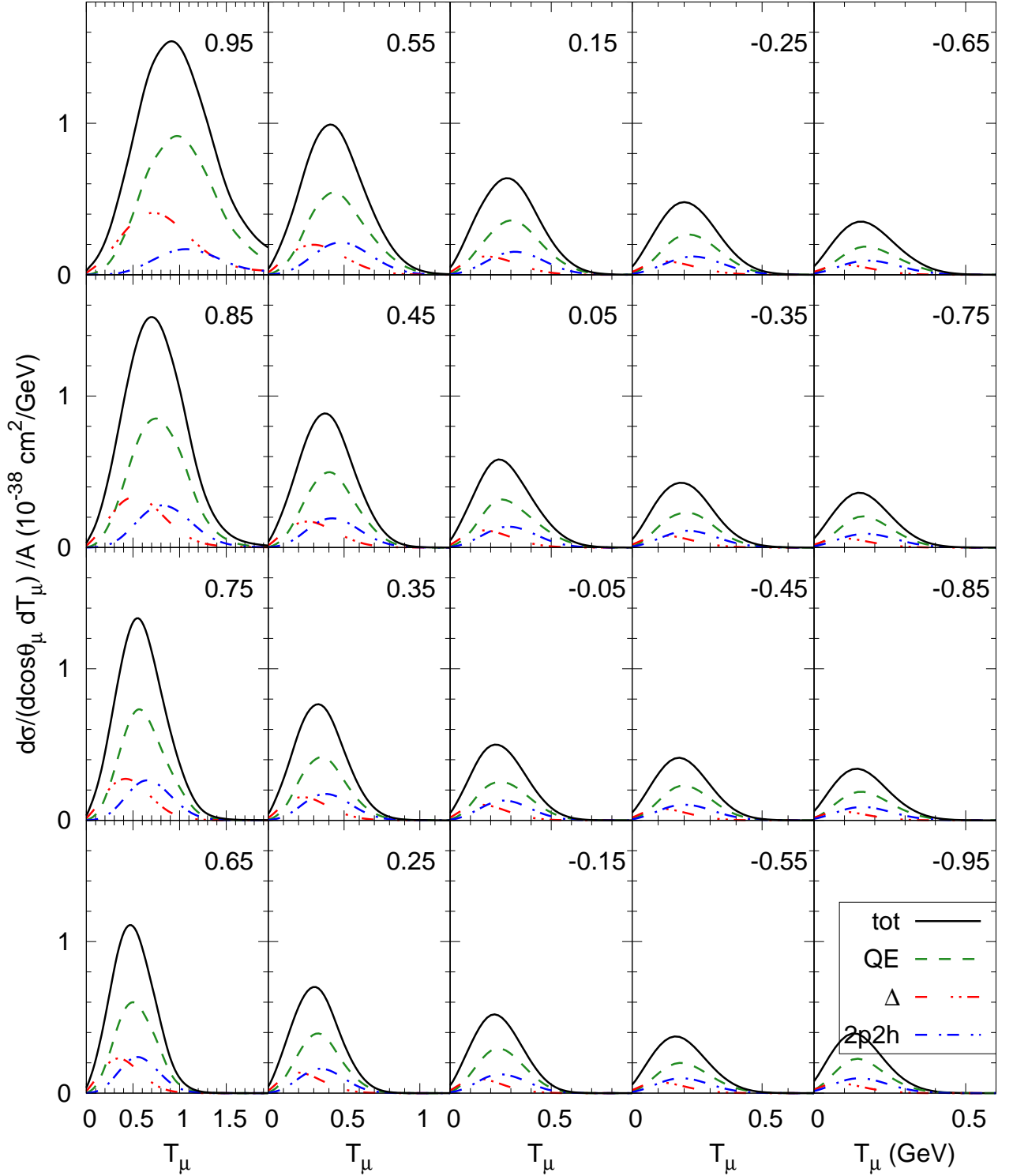}
	\caption{Fully inclusive double differential cross section per nucleon for the muon neutrino beam in the BNB at the MicroBooNE. The numbers in the upper right corner give the cosine of the muon scattering angle while $T_\mu$ is the muon's kinetic energy in GeV. The total inclusive cross section is given by the black solid line, the QE contribution by the green dashed line, the $\Delta$ contribution by the dash-dot-dotted line and the 2p-2h contribution by the blue dash-dotted line, as given in the figure. Other, smaller, contributions from pion-background and higher-resonance excitations are not explicitly shown in the figure, but are included in the total cross section. The target is $^{40}$Ar.}
	\label{fig:MicroB-nu}
\end{figure*}

The recently started experiment MicroBooNE \cite{Camilleri:2013oxa} works with the Booster Neutrino Beam (BNB) at Fermilab whose flux distribution is very similar to that of MiniBooNE. The difference is that now a heavier target, $^{40}$Ar, is used in a liquid argon detector. We have, therefore, also performed a calculation of the double-differential cross section for that target, using the BNB flux. The predicted cross section shown in Figure\ \ref{fig:MicroB-nu} is now, contrary to the one for MiniBooNE, not just a QE + 2p2p cross section, but instead a fully inclusive one.

It is seen that the $\Delta$ contribution is always as large or even larger (at forward angles) than the 2p2h contribution. This underlines the need to control this $\Delta$ contribution quantitatively if one is interested in a study of 2p2h processes. Tuning a generator such that just the total number of pions is reproduced is not sufficient to pin this contribution down. Instead, a double-differential cross section for the pions is necessary to make any analysis of 2p2h processes more quantitative.

Also in other aspects this double-differential distribution per nucleon does not differ significantly from the one obtained for the MiniBooNE. The higher target mass number mainly affects the fsi of outgoing particles while the initial interaction and thus the inclusive cross section per nucleon scales approximately with $A$. This is true even for the 2p2h interaction, if the interaction between the two nucleons is short-ranged (see discussion in Sect.\ \ref{A-dep} and \cite{Mosel:2016uge}). However, it is still a matter of ongoing debate whether the 2p2h correlations are indeed short ranged. A detailed experimental comparison of QE-like data on C (MiniBooNE) and Ar (MicroBooNE) could thus help to determine this property of the 2p2h interactions. MicroBooNE with its relatively low beam energy, and a flux that is very similar to that at MiniBooNE, should be ideally suited for that purpose since here QE and 2p2h constitute a major part of the total cross section.

\subsection{NOvA near detector results}
\begin{figure*}[h]
\centering
\includegraphics[width=0.8\linewidth]{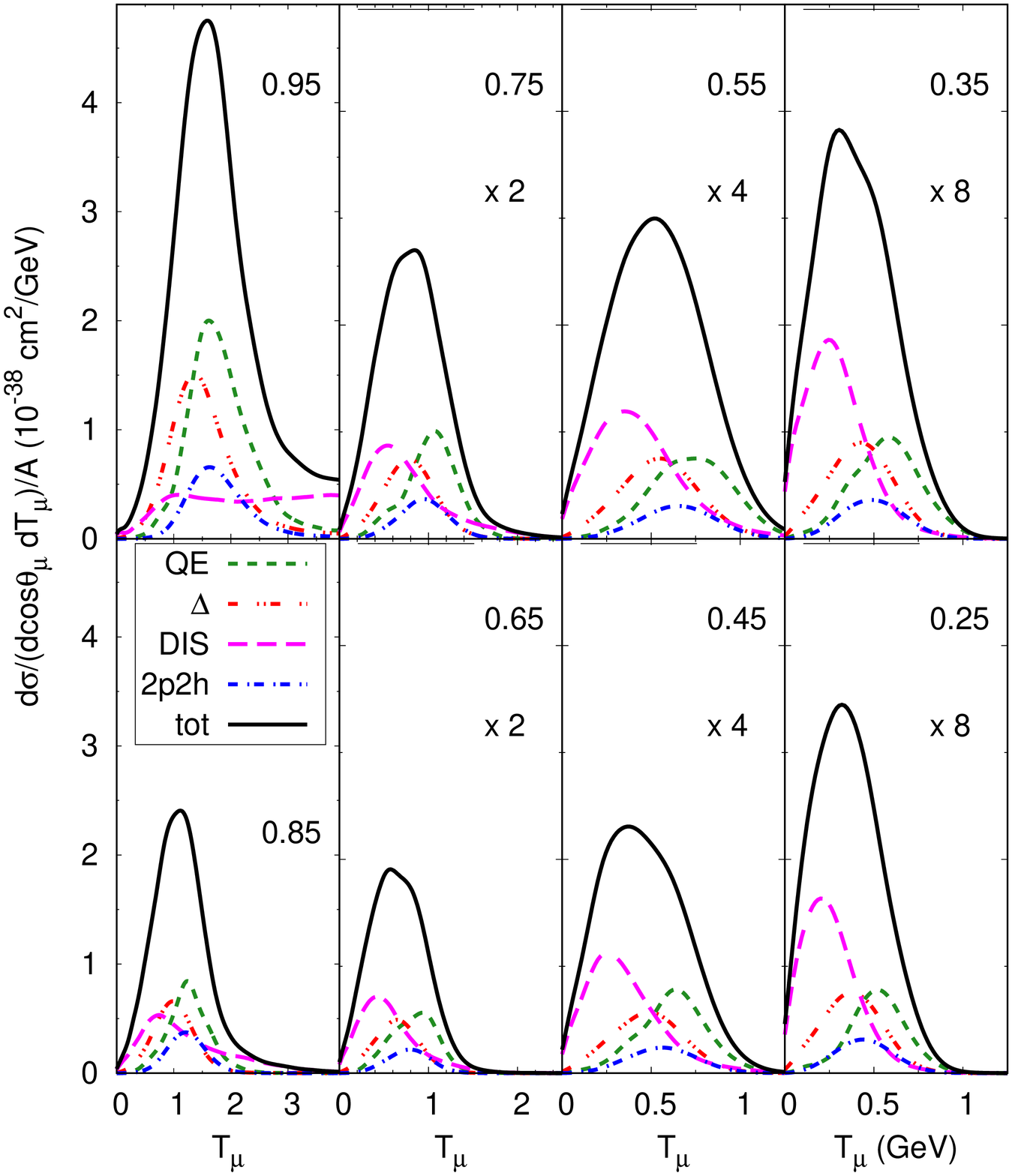}
\caption{Inclusive double differential cross section per nucleon for a C target with the muon neutrino beam in the NOvA near detector. The numbers in the upper right of each frame give the cosine of the scattering angle. The calculated curves for angles $\cos \theta \le 0.75$ have been scaled up by the factors given in the figures to enhance visibility. Curves as in Figure\ \ref{fig:MicroB-nu}, the additional dashed, magenta curve represents the DIS contribution.}
\label{fig:NOvA-nu-multiplot}
\end{figure*}

At higher energies the NOvA experiment works with a flux that is centered around 2 GeV. In Figure\ \ref{fig:NOvA-nu-multiplot} we show the  predicted inclusive double-differential cross section per nucleon for the muon neutrino flux at the NOvA near detector.
Immediately noticeable, in comparison to the results obtained for the lower energies at MiniBooNE, MicroBooNE and T2K, is the fact that the cross section is now much more forward-peaked. This can be understood by noting that the energy is higher and that most of the cross section still comes at rather small $Q^2 \approx 0.1 - 0.2$ GeV$^2$. The relation $Q^2 = 4 E_\nu E_\mu' \sin^2 \theta/2$ then leads to a dominance of small angles. Noticeable in the most forward bin ($\cos \theta = 0.95$) is the long-tailed, flat cross section at higher muon energies. This is caused  by DIS events that come in because even the NOvA flux extends up to high ($\approx $ 30 GeV) energies where DIS becomes dominant with $\sigma_{\rm DIS} \propto E_\nu$. For the larger angles with $\cos \theta \le 0.35$ DIS is the dominant component. It is connected with the largest energy loss and, therefore, peaks at the smallest $T_\mu$. For all angles true QE and $\Delta$ excitation are roughly equal in magnitude.

\subsection{NUMI beam results}
Compared to MiniBooNE and MicroBooNE a very different flux distribution is present in the NUMI beam. Neutrino energies reach up to much higher values and also the average energy lies significantly higher (approximately 3.5 GeV vs 0.7 GeV for T2K and MicroBooNE). The beam's energy profile is similar to that of the LBNF and the planned DUNE experiment. It is thus essential also for the future DUNE results to understand the reaction mechanisms in a quantitative way.

\begin{figure}[h]
\centering
\begin{minipage}{0.5\textwidth}
\includegraphics[width=1.\linewidth]{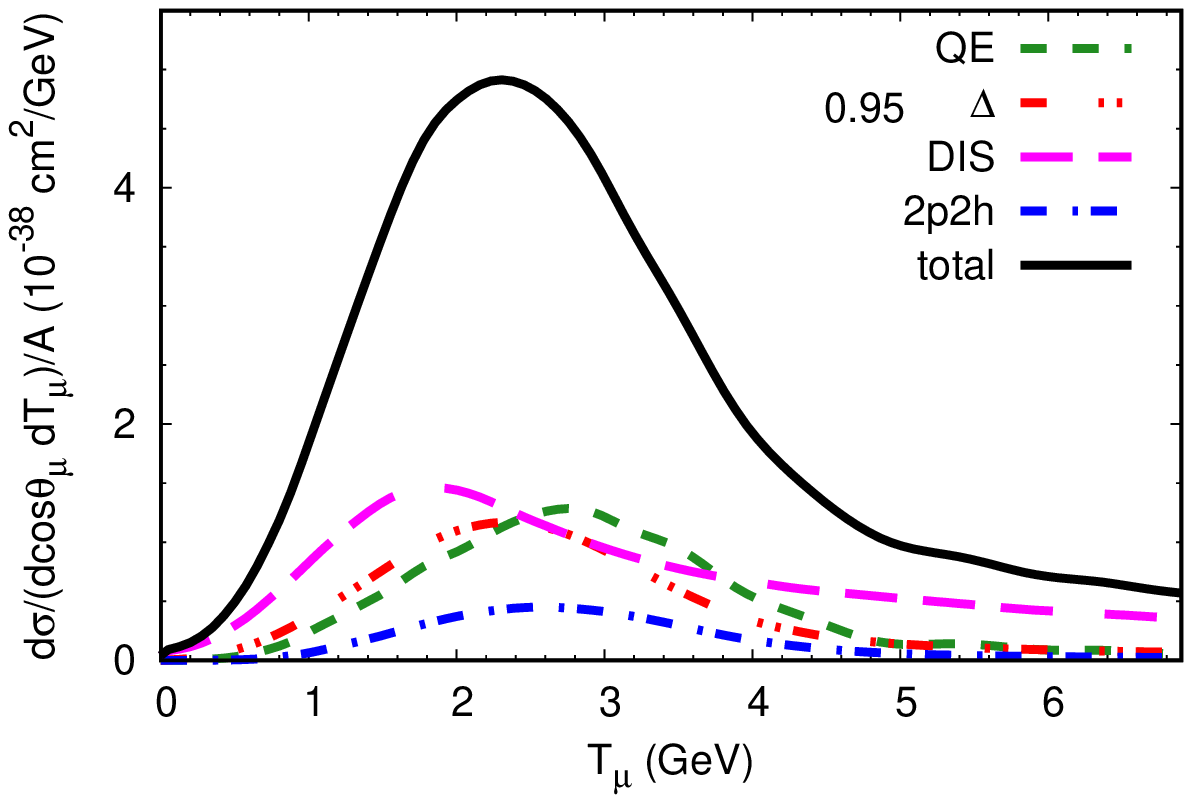}
\includegraphics[width=0.7\linewidth,angle=-90]{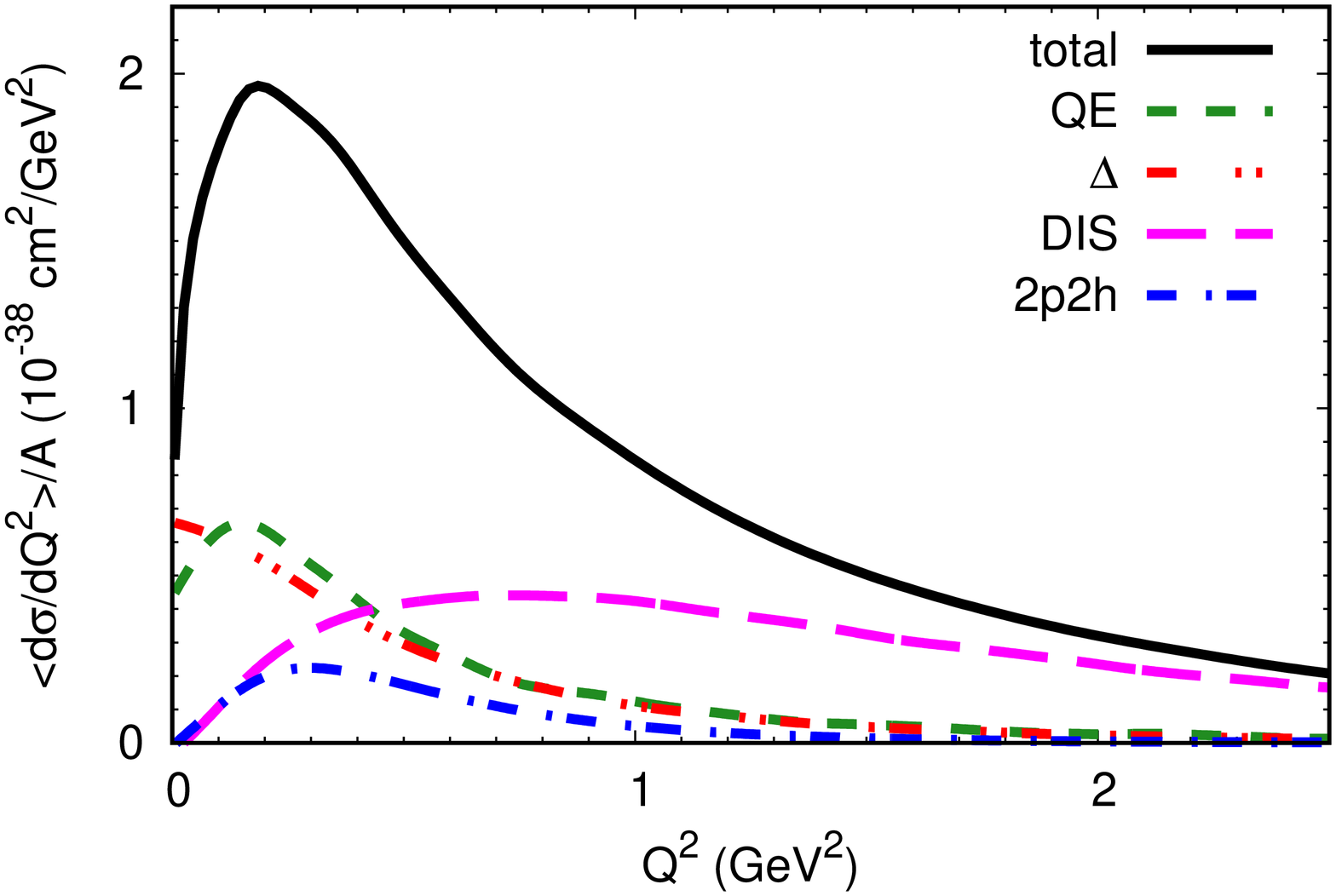}
\end{minipage}
\caption{(upper panel) Inclusive cross section per nucleon as a function of kinetic energy of the outgoing muon at the forward angle $\cos \theta = 0.95$ for the MINERvA experiment. (lower panel) Inclusive $Q^2$ distribution per nucleon for MINERvA. In both panels only the most relevant contributions are shown.}
\label{fig:MINERvA-nu}
\end{figure}

At present the experiment MINERvA operates in this beam. Its acceptance is such that only high-energy muons with scattering angles less than about 20 degrees make it into the muon detector. We, therefore, show in the upper part of Figure\ \ref{fig:MINERvA-nu} the inclusive cross section for a $^{12}$C target only for $\cos \theta = 0.95$, averaged over the MINERvA flux. For higher angles the cross section drops rapidly as already seen for the NOvA experiment, but even more pronounced here because of the still higher beam energy ($\langle E_\nu \rangle \approx 3.6$ GeV at MINERvA vs. 2 GeV for NOvA).

Thus all the neutrino-nucleus interaction physics at the MINERvA experiment is concentrated in a very forward direction. Furthermore, the muon spectrometer used in that experiment sees only events with muon energies above about 1.5 GeV, so that the left quarter in the upper panel in Figure\ \ref{fig:MINERvA-nu} is blocked out and the reaction dynamics have to be reconstructed from the visible remainder. Both of these facts make the necessary energy reconstruction more difficult \cite{Mosel:2014lja}. They may also -- at least partly -- explain the difficulty to extract a convincing 2p2h signal from these data \cite{Fiorentini:2013ezn,Walton:2014esl}. This is also complicated by the fact
that the 2p2h cross section comes up to only about 1/3 of the $\Delta$ and true QE contributions. To isolate QE and the relatively small 2p2h cross sections from this multitude of other, stronger reaction mechanisms, including DIS, requires considerable generator work, with the danger of 'generator contamination' of the data. The latter is minimized in the inclusive cross sections discussed here.

The $Q^2$ distribution exhibits the usual behavior with a peak around $Q^2 \approx 0.15$ GeV$^2$ (lower part of Figure\ \ref{fig:MINERvA-nu}). $\Delta$- and DIS-contributions together are significantly larger than the sum of QE and 2p2h reactions. At $Q^2 \gtrsim 0.5$ GeV$^2$ DIS dominates. Thus, MINERvA is an ideal experiment to explore pion production \cite{Eberly:2014mra,Aliaga:2015wva} and DIS \cite{Tice:2014pgu,Mousseau:2016snl} (which mostly contributes also to pion production), the two dominant components of the cross section. In particular our understanding of the theoretically difficult transition from the resonance-dominated to the DIS-dominated energy region could be improved significantly in experiments such as NOvA or MINERvA. For a more detailed discussion of reaction mechanisms at MINERvA we refer the reader to \cite{Mosel:2014lja}.

\subsection{LBNF-DUNE Results}

\begin{figure*}[ht]
\centering
\includegraphics[width=\linewidth]{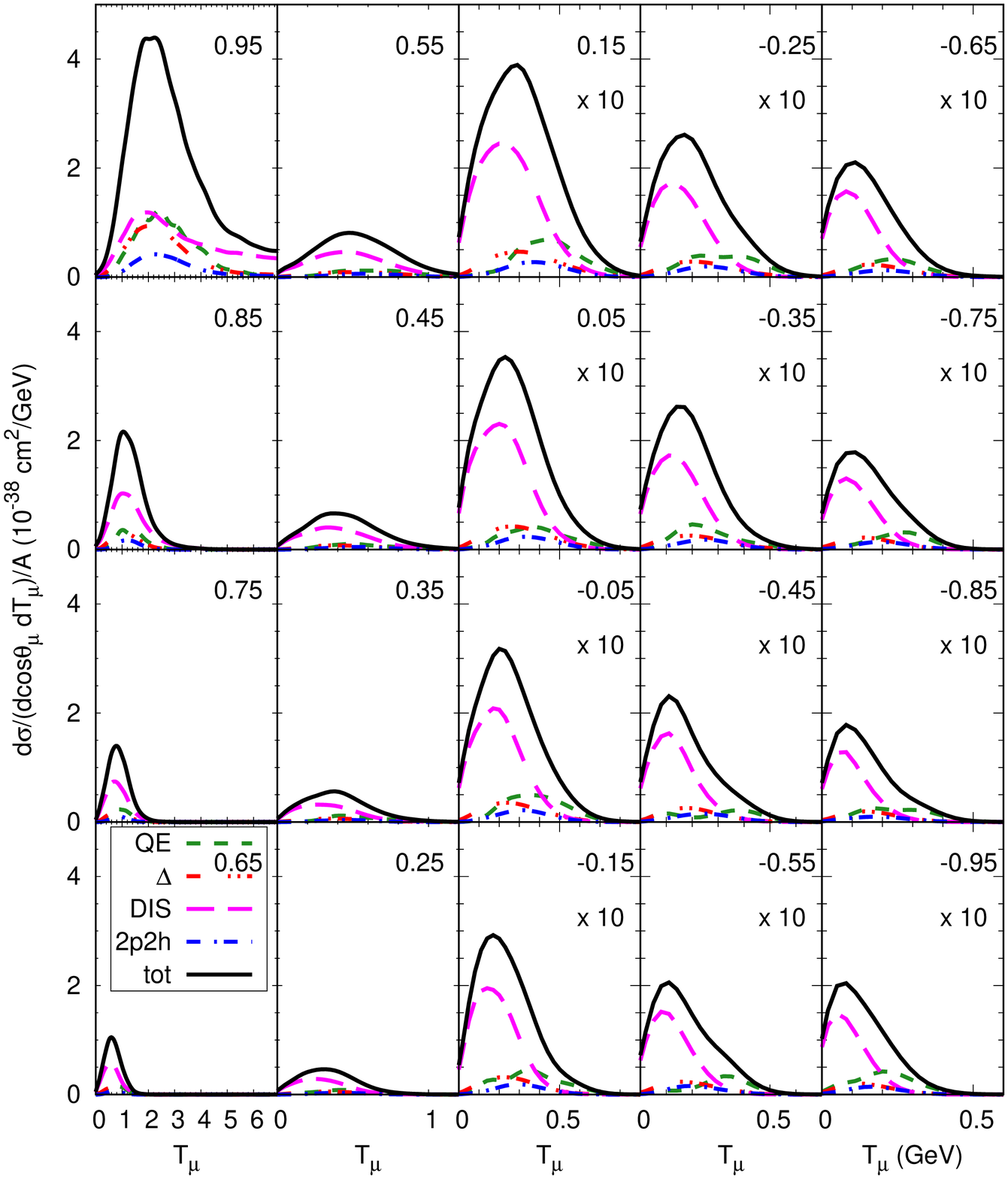}
\caption{Fully inclusive double differential cross section per nucleon for the muon neutrino beam from the LBNF. The numbers in the upper right corner give the cosine of the muon scattering angle while $T_\mu$ is the muon's kinetic energy in GeV. For angles with $\cos \theta \le 0.15$ the cross section scale has been changed by a factor of 10 to make the composition at backward angles more visible. The dominant reaction contributions are denoted as given in the figure.}
\label{fig:DUNE-nu}
\end{figure*}
The beam at the planned LBNF-DUNE experiment is in its flux similar to that of NUMI. Since the planned DUNE detector will have a $4\pi$ coverage in a liquid Argon chamber, it is interesting to investigate here also the cross sections under backward angles in some more detail. In Figure\ \ref{fig:DUNE-nu} we show the inclusive double-differential cross section expected at LBNF (or DUNE, without oscillation), broken up into its most important components.
At forward angles true QE, DIS and $\Delta$ excitation are all of about the same magnitude this changes rapidly with increasing angle. Already for $\cos \theta = 0.85$ DIS is the largest component and it becomes dominant for even larger or backward angles. The other components are all significantly smaller and similar to each other in their absolute magnitude. 2p2h processes are even smaller and play no significant role.

\subsection{Open Problems}
\subsubsection{Magnitude of 2p2h effects}   \label{s:2p2hsize}
The discussions in the preceding sections have shown that the relative importance of the 2p2h contribution to the total cross section decreases with increasing beam energy simply because inelastic channels pick up more and more strength.

It should be noted that all the calculations were performed assuming an isospin $\mathcal{T} = 1$ in Eq.\ (\ref{sigma-nu}). The good agreement reached in particular with the MiniBooNE data seems to indicate that mostly the $\mathcal{T}=1$ pairs contribute to the neutrino 2p2h interactions. While most of the short-range nucleon-nucleon pairs seem to be in $\mathcal{T}=0$ states in the nucleus \cite{Colle:2015ena}, it is not clear if this preference also survives once the coupling to the incoming neutrinos is taken into account. A detailed comparison of calculations with experiment could thus help to clarify this question. However, unfortunately, the accuracy of presently available data does not allow for such a clarification. Using in all the calculations above $\mathcal{T}=0$ would have cut the 2p2h contributions  by a factor of 2. The total cross section, however, would be much less affected, by only about 10\%. A closer inspection of all the figures shows that these calculations with $\mathcal{T}=0$ would then still fit the fully inclusive T2K data. For MiniBooNE agreement could again be restored by changing the data because of flux uncertainties which also amount to about 10\%. This latter change was, for example, exploited in the comparison of calculated results with data by Nieves \textit{et al.}\ \cite{Nieves:2011yp}. Presently available data thus do not allow to determine the neutrino-induced 2p2h processes to better than within a factor of 2. For this situation to change the flux would have to be known to significantly better than 10\%.

\subsubsection{Structure functions at large energy transfers}
The data analyzed by Bosted \textit{et al.}\ \cite{Bosted:2012qc} included no photoabsorption data. From these it is known that there are 2p2h1$\pi$ contributions to the total absorption cross section at $\omega \gtrsim 300$ MeV that set in on the high-energy side of the $\Delta$ resonance \cite{Carrasco:1989vq,Effenberger:1996im,Gil:1997bm}. Since these processes will also be present at $Q^2 > 0$ they would fill in the high-energytransfer part of $W_1$ and may thus have some effects in the higher-energy experiments such as MINERvA and DUNE. These additional reaction components amount to about 10 - 20\% of the total photoabsorption cross section \cite{Gil:1997bm} (see also Figure\ 9 in \cite{Effenberger:1996im}) and we expect them to have an approximately equal magnitude also for neutrinos. However, none of the existing microscopic model calculations includes such 2p2h1$\pi$ terms for neutrinos. This is a problem for future theoretical work.

\section{Summary and Outlook}
In this article we have shown that it is possible to obtain a reliable theoretical description of electron-~, neutrino- and antineutrino-nucleus data over a wide range of energies and different experiments without any special tuning. The description is based on the GiBUU model of lepton-nucleus interactions which uses consistently the same ground state properties for all reaction processes. It is also, by construction, consistent with electron-nucleus experiments and has been checked against these. While the model includes a quantum-kinetic transport-theoretical description of final state interactions these have not explicitly been used in the present study which was concentrated on inclusive cross sections. Implicitly, however, these same fsi play a role also in the determination of the final state of the initial neutrino-nucleon interaction and there they have been included.

The two major improvements to GiBUU that have made this wide-ranging description possible are, first, a better description of the nuclear ground state that now has an increased stability compared to earlier calculations within this model. This plays a role mainly at low momentum-transfers. Second, the description of 2p2h interactions has now significantly been improved by using the 2p2h structure function obtained from an analysis of electron scattering data. By assuming at the same time a purely transverse character of these interactions it has been possible to base all the electron, neutrino and antineutrino cross sections on this one structure function. All the available data for inclusive differential cross sections are well described within this framework.

In addition to these inclusive cross sections GiBUU also generates full event files with four-vectors for all final-state particles and information on their history; it can thus be used as a generator. GiBUU is thus a theory-framework and tool that can be applied to the analysis of ongoing and planned neutrino long-baseline experiments. Since it does not involve any tuning to neutrino data, it has predictive power for new energy regimes or new targets.

Open theoretical problems are, first, the behavior of the structure functions at higher energy transfers. Here, evidence from photo- and electro-absorption measurements indicates that 2p2h processes become important that contain $\Delta$ or higher $N^*$ excitations in the final state.  Since these processes will also be present at $Q^2 > 0$ they could contribute to the high-energytransfer part of $W_1$. Such excitations have so far not been considered for neutrino interactions. To assess their actual, quantitative importance requires some theoretical work that generalizes the earlier work on photoabsorption to the electroweak sector. It is interesting to note that such 2p2h processes with an outgoing pion from resonance decay could also increase the pion yield through 2p2h$1\pi$ production processes.

The second open problem is how to consistently embed the 2p2h processes into a full event generation without possible double counting. Using only inclusive information does not give any information on the final state momenta and, in particular, on their angular distribution. So far, practical implementations of 2p2h processes in generators (including GiBUU) all use an isotropic phase-space distribution. It is also not clear how to correctly include the 2p2h diagrams with an intermediate $\Delta$ in actual event generators since these usually already contain processes such as $\Delta N \to NN$ in their fsi. This also deserves some future study.

A third open problem is the isospin content of the two-nucleon pairs in 2p2h interactions since this directly affects the magnitude of the related cross sections. While nuclear structure calculations seem to give a preference for $\mathcal{T}=0$ pairs due to spin-isospin interactions in nuclear matter, the actual coupling of neutrinos to these pairs may well prefer $\mathcal{T}=1$. 

Finally, an interesting question is whether the 2p2h interactions are short- or long-ranged; the latter possibility receives some support from the observation that the interactions are caused by meson exchange. This should show up in the $A$-dependence of 2p2h cross sections. For zero-range interactions the mass-number dependence should go $\propto A$ (with modifications due to the nuclear surface at small $A$), whereas for the extreme of long-range interactions it should go $\propto A^2$. Ideal for an investigation of this question are experiments at the lower energies, e.g., at the T2K near detector or the MicroBooNE. There the pion background does not yet play the major role so that any generator-dependence of this background subtraction is minimized.

\begin{acknowledgments}
We gratefully acknowledge the help and support of the whole GiBUU team in developing both the physics and the code used here. One of the authors (U.M.) also is grateful for helpful discussions with D. Day and M. Martini. He also thanks E.\ Christy for providing him with the code necessary to determine the MEC structure function from electron scattering data.

This work has partially been supported by DFG.
\end{acknowledgments}

	
\end{document}